\documentclass[a4paper,amsfonts,notitlepage,11pt,superscriptaddress,nofootinbib,floatfix,showpacs]{revtex4-1}
\pdfoutput=1
\usepackage[T1]{fontenc}
\usepackage[utf8]{inputenc}
\usepackage{amsmath,amssymb,amsfonts}
\usepackage{mathrsfs}
\usepackage{bbm}
\usepackage{slashed}
\usepackage[dvipdf,dvips,dvipdfmx]{graphicx}
\usepackage{verbatim}

\usepackage[english]{babel}
\usepackage{mathbbol}
\usepackage{aas_macros}
\usepackage[colorlinks]{hyperref}

\def\be{\begin{equation}}
\def\ee{\end{equation}}
\def\ba{\arraycolsep .1em \begin{eqnarray}}
\def\ea{\end{eqnarray}}

\def\apjl{ApJ}%
\def\apjs{ApJS}%
\def\ao{Appl.\ Opt.}%
\def\aap{A\&A}%
\def\jcap{JCAP}%
\def\prd{Phys.\ Rev.\ D}%
\def\physrep{Phys.\ Rep.}%

\begin{document}

\title{The inflationary mechanism in Asymptotically Safe Gravity}

\author{Alessia Platania}
\email{a.platania@thphys.uni-heidelberg.de}
\affiliation{Institut für Theoretische Physik, Universität Heidelberg, Philosophenweg 16, 69120 Heidelberg, Germany}

\begin{abstract}
{According to the asymptotic safety conjecture, gravity is a renormalizable quantum field theory whose continuum limit is defined by an interacting fixed point of the renormalization group flow. In these proceedings we review some implications of the existence of this non-trivial fixed point in cosmological contexts. 
Specifically, we discuss a toy model exemplifying how the departure from the fixed-point regime can explain the approximate scale-invariance of the power spectrum of temperature fluctuations in the cosmic microwave background.}
\end{abstract}

\maketitle

\section{Introduction}

Primordial quantum fluctuations occurring in the pre-inflationary
epoch have left indelible imprints which we measure today in the form of tiny temperature anisotropies, $\delta T/T\sim10^{-5}$, in
the Cosmic Microwave Background (CMB) radiation. 
The inflationary mechanism furnishes a simple explanation for the presence of these anisotropies \cite{Baumann:2009ds} and it has become a paradigm in the description of the primordial evolution of the universe within the standard cosmological model. 

The spectrum of the CMB reproduces an almost perfect black-body radiation at an average temperature $\langle T\rangle\sim2.7K$.
The distribution of temperature fluctuations in the CMB is described by the power spectra of scalar and tensorial perturbations. 
These spectra are essentially characterized by two parameters: the spectral index $n_s$, giving information on the scale dependence of the power spectrum of scalar fluctuations, and the tensor-to-scalar ratio $r$, measuring the suppression of tensorial perturbations against the scalar ones. 
The values of the spectral index $n_{s}$ and tensor-to-scalar ratio $r$ can be obtained from the observational data. In particular, the most recent observations to date \cite{Akrami:2018odb} constrain the spectral index to be $n_{s}=0.9649\pm0.0042$ at $68\%$ CF, and limit the tensor-to-scalar ratio to values $r<0.064$. Note that, although the scalar power spectrum is almost scale invariant, perfect scale invariance, corresponding to $n_s=1$, seems to be excluded. 

The extraordinary predictive power of the inflationary scenario, combined with the current limits on the determination of $r$, makes it difficult to distinguish between different models of cosmic inflation \cite{inflationaris}. The simplest inflationary model capable of explaining the current observational data is the Starobinsky model \cite{1980staro}. In the Einstein frame the only free parameter of the model is the inflaton mass, and this mass is fixed by the normalization of the amplitude of the scalar power spectrum \cite{Akrami:2018odb}. In addition, the Starobinsky Lagrangian is conformally equivalent to Einstein gravity coupled to the Standard-Model Higgs-boson by means of the non-minimal interaction term~$\xi H^{\dagger}HR$~\cite{Birrell:1982ix,Bezrukov:2007ep}, making this model particularly interesting.

Although Starobinsky inflation and other models characterized by an inflationary potential with a plateau are favored by observations \cite{Akrami:2018odb}, it has been argued that they might re-introduce the fine-tuning problems that inflation is supposed to solve, resulting in the so-called ``unlikeness problem'' \cite{Ijjas:2013vea}. In fact, the upper bound on the tensor-to-scalar ratio arising from the CMB data lowers the scale of inflation down to $\sim10^{16} \mathrm{GeV}$, and a proper resolution of the flatness and horizon problems requires an inflationary potential with $V_{plateau}(\phi)\sim M_{Pl}^4$ \cite{Ijjas:2013vea}. A more fundamental understanding of inflation and its role in the cosmological evolution of the universe thus requires these inflationary models to be understood and embedded in a more general framework, explaining the origin of the plateau based on first principles, e.g. on short-distance modifications of General Relativity due to quantum gravity.

Adopting the Wilsonian point of view, a quantum field theory is well defined and predictive if its renormalization group flow is equipped with an ultraviolet fixed point, endowed with a finite-dimensional basin of attraction. This fixed point, ensuring the renormalizability of the theory~\cite{Wilson:1973jj}, can be Gaussian (GFP), corresponding to a free theory, or non-Gassian (NGFP). In the latter case the fundamental theory is interacting, and termed ``asymptotically safe''. As shown by numerous computations \cite{Reuter:1996cp,Souma:1999at,Reuter:2001ag,Litim:2003vp,Codello:2007bd,Benedetti:2009gn,Groh:2011vn,Donkin:2012ud,Benedetti:2013jk,Eichhorn:2013xr,Falls:2014tra,Demmel:2015oqa,Eichhorn:2015bna,Biemans:2016rvp,Gies:2016con,Hamada:2017rvn,Biemans:2017zca,Platania:2017djo,Falls:2018ylp} employing functional renormalization group (FRG) techniques \cite{Polonyi:2001se,eaa}, the gravitational RG flow could attain a NGFP in the ultraviolet limit. The asymptotic-safety mechanism would hence allow to quantize gravity in the well-estabilished framework of quantum field theory.

In these proceedings we exploit the consequences of the existence of a gravitational fixed point in cosmological contexts. To this end we will use the toy model constructed in \cite{Bonanno:2018gck}, where an effective action for inflation was derived from the renormalization group improvement of the Einstein-Hilbert action (see also~\cite{irfp,resa05,weinberg10,alfio12,copeland,AlfioAle1,Bonanno:2016rpx}). Using the equivalence of the description in the Jordan and Einstein frames, it will be shown explicitly that a period of slow-roll inflation can be generated by the departure of the RG flow from the scale-invariant regime associated with the cosmological ``fixed-point era''~\cite{irfp,br02}. We will show that this scenario has two important consequences. First of all, the generation of a nearly-scale-invariant power spectrum of scalar perturbations in the CMB can be seen as a relict of the nearly-scale-invariant behavior of the gravitational RG flow in the vicinity of its ultraviolet attractor. In this scenario the inflationary potential is characterized by a plateau region; however, due to the running of the gravitational couplings, the bound on the tensor-to-scalar ratio might not necessarily re-introduce the fine-tuning problems discussed in \cite{Ijjas:2013vea}. Secondly, the inflationary potential depends on universality properties of the gravitational interaction and therefore compatibility with observations can constrain the way the RG trajectories depart from the fixed point and put bounds on the corresponding critical exponents \cite{Bonanno:2018gck}.

This paper is organized as follows. In sect. \ref{sect2} we review key features of Asymptotically Safe Gravity and introduce a variant of the RG-improved model constructed in \cite{Bonanno:2018gck}. Sect. \ref{sect3} discusses the conformal representation of RG-improved theories and their relation to observations. Assuming that the dynamics and output of cosmic inflation are determined solely by the quantum fluctuations of the spacetime, in sect. \ref{sect4}  we provide a simplified but explicit explanation of how a period of slow-roll inflation can be triggered by the departure of the RG flow from the scale-invariant regime defined by the interacting fixed point. Finally, sect. \ref{sect5} summarizes  our conclusions.

\section{Running couplings and effective actions} \label{sect2}

In the $(G,\Lambda)$-theory-space, the beta functions for the dimensionless Newton coupling~$g_{k}=G_{k}k^{d-2}$ and cosmological constant $\lambda_{k}=\Lambda_{k}k^{2}$ are determined by the projection of the functional renormalization group equation (FRGE) \cite{eaa,Morris:1993qb,ReuterWetterich} on the Einstein-Hilbert subspace. The renormalization group equations
for $g_{k}$ and $\lambda_{k}$ can generally be written as
\begin{equation}
\begin{cases}
k\partial g_{k}=\beta_{g}(g_{k},\lambda_{k})=\left\{ d-2+\eta_{G}(g_{k},\lambda_{k})\right\} g_{k}\\
k\partial\lambda_{k}=\beta_{\lambda}(g_{k},\lambda_{k})
\end{cases}\;,
\end{equation}
where $\eta_{G}\equiv\frac{\partial\log G_{k}}{\partial\log k}$ is
the anomalous dimension of the Newton coupling. A non-trivial fixed
point~$(g_{\ast},\lambda_{\ast})$ exists if~$\beta_{\lambda}(g_{\ast},\lambda_{\ast})$ vanishes and at the same time the anomalous dimension $\eta_{G}$ flows to~$\eta_{G}(g_{\ast},\lambda_{\ast})=2-d$. The existence of a non-gaussian fixed point (NGFP) thus entails an effective dimensional reduction from~$d=4$ to~$d_{\text{{eff}}}=2$ spacetime dimensions \cite{Lauscher:2005qz,Reuter:2011ah}.
This property seems to be a common prediction of several approaches to quantum gravity~\cite{2005Loll,2009horava,2009modesto,2014Amelino}.
A key consequence of this dimensional reduction is a modification
of the graviton propagator at short distances. Under certain approximations, it scales as~$\mathcal{G}(x,y)\sim\log|x-y|^{2}$ for $\eta=-2$ \cite{Lauscher:2005qz} and gives rise to a scale-invariant scalar power spectrum~\cite{Lauscher:2002sq,br02,Lauscher:2005qz,br07}.
It is thereby possible that the nearly scale-invariant power spectrum of temperature fluctuations in the CMB arises from the nearly scale-invariant regime following the NGFP epoch. In order to understand if and under which conditions this mechanism is realized, it is important to study the departure of the renormalization group (RG) flow from its ultraviolet fixed point.

The universality properties of the RG flow about
the non-trivial fixed point $(g_{\ast},\lambda_{\ast})$ are determined by the stability matrix $\partial_{g_{i}}\beta_{j}(g)|_{g_{\ast}}$. Denoting by $\textbf{e}_{i}$ its eigenvectors and by $(-\theta_{i})$ the corresponding eigenvalues, the running of the dimensionless gravitational couplings about~$(g_{\ast},\lambda_{\ast})$ can be written as
\begin{equation}
\begin{cases}
g_{k}= & g_{\ast}+c_{1}e_{1}^{1}\left(\frac{k}{M_{Pl}}\right)^{-\theta_{1}}+c_{2}e_{2}^{1}\left(\frac{k}{M_{Pl}}\right)^{-\theta_{2}}\\
\lambda_{k}= & \lambda_{\ast}+c_{1}e_{1}^{2}\left(\frac{k}{M_{Pl}}\right)^{-\theta_{1}}+c_{2}e_{2}^{2}\left(\frac{k}{M_{Pl}}\right)^{-\theta_{2}}
\end{cases}\;.\label{eq:scalingUV}
\end{equation}
where $M_{Pl}\sim(8\pi G_{0})^{-1/2}$ is the reduced Planck mass.

The NGFP typically found in the functional renormalization group (FRG) computations in the Einstein-Hilbert truncation is characterized by $\mathrm{Re}(\theta_{1})>0$ and $\mathrm{Re}(\theta_{2})>0$. 
The positivity of the real part of the universal critical exponents
$\theta_{i}$ indicates that, in the aforementioned truncation, the
fixed point $(g_{\ast},\lambda_{\ast})$ is endowed with two relevant directions. FRG computations in higher-order truncations show that there might be one more relevant direction, associated with 4th-order derivative operators. These relevant directions identify the UV critical surface. The NGFP can thus act as an ultraviolet ``sink'' for the RG trajectories belonging to its basin of attraction. The constants $c_{i}$ are integration constants, corresponding to different initial conditions of the flow. Every pair $(c_{1},c_{2})$ identifies a particular RG trajectory. These free parameters should be fixed by equating the infrared values of the dimensionful running couplings $G_{k}$ and $\Lambda_{k}$ with the values of the Newton and cosmological constants at observational scales, namely $8\pi G_{0}\sim M_{Pl}^{-2}$ and $\Lambda_{0}\sim3\cdot10^{-122}M_{Pl}^{2}$. This comparison allows to select the particular RG trajectory realized by Nature \cite{Reuter:2004nx}.

Starting from a classical action of the form
\begin{equation}
S_{cl}=\frac{1}{16\pi G_{0}}\int d^{4}x\sqrt{-g}\;(R-2\Lambda_{0})+S_{matter}\;,\label{eq:claction}
\end{equation}
the introduction of quantum effects typically results in the emergence of higher-derivative terms. The renormalization effects thus modify the
interactions of the theory and, as a consequence, the coupling constants appearing in the bare Lagrangian turn into running functions of the energy (or length) scale. Reversing the argument, replacing the coupling constants in the classical action with running functions and promoting the corresponding energy-scale to a proper coordinate-dependent quantity~$k=k(x)$ should provide an effective action which mimics, at least qualitatively, the effects of quantum loops~\cite{Polonyi:2001se,2004reuterw2}. Neglecting the running of the matter couplings, the scale-dependent version of
the action~$\eqref{eq:claction}$ is
\begin{equation}
S_{k}=\frac{1}{16\pi G_{k}}\int d^{4}x\sqrt{-g}\;(R-2\Lambda_{k})+S_{matter}\;,\label{eq:qaction}
\end{equation}
and the corresponding field equations read \cite{2004reuterw2}
\begin{equation}
G_{\mu\nu}=8\pi G_{k}T_{\mu\nu}+\Lambda_{k}g_{\mu\nu}+\Delta t_{\mu\nu}\;.
\end{equation}
Here $G_{\mu\nu}$ is the Einstein tensor, $T_{\mu\nu}=-\frac{2}{\sqrt{-g}}\frac{\delta S_{matter}}{\delta g^{\mu\nu}}$, 
and $\Delta t_{\mu\nu}\equiv G_{k}(\nabla_{\mu}\nabla_{\nu}-g_{\mu\nu}\square)G_{k}^{-1}$ is an effective energy-momentum tensor generated by the running of the Newton coupling \cite{2004reuterw2}. 

Assuming that there is no energy-momentum flow between the gravitational and matter components of the theory, i.e. that the energy-momentum tensor $T_{\mu\nu}$ is separately conserved, the momentum-scale $k=k(x)$ is determined by a set of of consistency conditions dictated by the Bianchi identities \cite{2004reuterw1,2004reuterw2,Babic:2004ev,Domazet:2012tw,Koch:2014joa}. In particular, diffeomorphism invariance requires \cite{2004reuterw1,Domazet:2012tw,Koch:2014joa}
\begin{equation}
\partial_kG_{k}\,R=2(\Lambda'_{k}G_{k}-G'_{k}\Lambda_{k})\;.\label{eq:constraint}
\end{equation}
In the fixed point regime, the scaling of the dimensionful Newton
coupling and cosmological constant read
\begin{equation}
G(k)=g_{\ast}k^{-2}\;,\qquad\Lambda(k)=\lambda_{\ast}k^{2}\;\;,\label{eq:scaling}
\end{equation}
Combining eq. $\eqref{eq:scaling}$ with the constraint $\eqref{eq:constraint}$ yields \cite{Domazet:2012tw,Koch:2014joa}
\begin{equation}
k^{2}=\frac{R}{4\lambda_{\ast}}\;\;.\label{eq:cutid}
\end{equation}
A similar relation should also hold when additional operators of the form $R^n$ are added to the bare Lagrangian, at least in the fixed-point regime \cite{Domazet:2012tw}.

The replacement $k^{2}\to R/4\lambda_\ast$ in the scale-dependent action $\eqref{eq:qaction}$ generates an effective
$f(R)$ action, whose analytical expression is determined by the running of the gravitational couplings \cite{alfio12,Domazet:2012tw,saltas12}.
In particular, in the vicinity of the NGFP, the running $\eqref{eq:scalingUV}$ leads to the following effective action \cite{Bonanno:2018gck}
\begin{equation}
S_{grav}^{\text{{eff}}}=S_{grav}^{\ast}+\int d^{4}x\sqrt{-g}\;\left(b_{1}R^{\frac{4-\theta_{1}-\theta_{2}}{2}}+b_{2}R^{\frac{4-\theta_{1}}{2}}+b_{3}R^{\frac{4-\theta_{2}}{2}}+b_{4}R^{2-\theta_{1}}+b_{5}R^{2-\theta_{2}}\right)\;,\label{eq:effaction}
\end{equation}
where the fixed-point action is given by \cite{Domazet:2012tw}
\begin{equation}
S_{grav}^{\ast}=\int d^{4}x\sqrt{-g}\;\frac{R^{2}}{128\pi g_{\ast}\lambda_{\ast}}\;\;,\label{eq:FPaction}
\end{equation}
and the coefficients $b_{i}$ read \cite{Bonanno:2018gck}
\begin{subequations}
\begin{align}
&b_1=\frac{c_1 c_2\, (e_1^1 \,e_2^2{+e_1^2 \,e_2^1})\,(4\lambda_\ast M_{Pl}^2)^{\frac{\theta_1+\theta_2}{2}}}{128\pi (g_\ast\lambda_\ast)^2} \;\;, \\
&b_2=\frac{c_1\, ( e_1^2\,\lambda_\ast-e_1^1\, g_\ast -2e_1^2\,\lambda_\ast)\,(4\lambda_\ast M_{Pl}^2)^{\frac{\theta_1}{2}}}{128 \pi  (g_\ast\lambda_\ast)^2} \;\;, \\
&b_3=\frac{c_2\, ( e_2^2\,\lambda_\ast-e_2^1 \, g_\ast -2e_2^2\,\lambda_\ast)\,(4\lambda_\ast M_{Pl}^2)^{\frac{\theta_2}{2}}}{128 \pi  (g_\ast\lambda_\ast)^2}\;\;, \\
&b_4= \frac{c_1^2\, (e_1^1 \, e_1^2)\,(4\lambda_\ast M_{Pl}^{2})^{{\theta_1}}}{128\pi (g_\ast\lambda_\ast)^2}\;\;, \\
&b_5=\frac{c_2^2\, (e_2^1 \, e_2^2)\,(4\lambda_\ast M_{Pl}^{2})^{{\theta_2}}}{128\pi (g_\ast\lambda_\ast)^2} \;\;.
\end{align}
\end{subequations}
In what follows the critical exponents $\theta_i$ will be assumed to be real numbers, as indicated by computations of the gravitational RG flow in the presence of matter fields \cite{Biemans:2017zca,Alkofer:2018fxj}.

The combination of the scaling relation~\eqref{eq:cutid} with the scale-dependent Einstein-Hilbert action \eqref{eq:qaction} correctly reproduces the fixed-point Lagrangian $f_{\ast}(R)\sim R^{2}$ found in the study of the renormalization group flow of $f_{k}(R)$-gravity theories \cite{Benedetti:2012dx,Dietz:2012ic,Demmel:2015oqa}. Notably in cosmological contexts the action $f_{\ast}(R)\sim R^{2}$ gives rise to a perfectly scale-invariant power spectrum, $n_{s}=1$ (flat inflationary potential).
By lowering the energy-scale $k^{2}\sim R$ towards the infrared,
the gravitational RG flow departs from the fixed-point regime and
generates additional operators in the Lagrangian, $\mathcal{L}_{grav}^{\text{{eff}}}=\mathcal{L}_{\ast}+\delta\mathcal{L}_{RG}$. As is clear from eq.~$\eqref{eq:effaction}$, the form of the deviation~$\delta\mathcal{L}_{RG}\equiv f_{RG}(R)$ depends crucially on the running of the gravitational couplings. Therefore, the study of the inflationary scenario arising from the RG-improvement of the Einstein-Hilbert action might actually put constraints on microscopic details of the theory, as for instance its critical exponents~\cite{Bonanno:2016rpx,Bonanno:2018gck}. Moreover, the nearly-scale invariant power spectrum of temperature anisotropies in the CMB might be related to the deviation of the Lagrangian~$\mathcal{L}_{grav}^{\text{{eff}}}$ from the scale-invariant regime described by~$\mathcal{L}_{\ast}$. This will be shown explicitely in the following sections. 

\section{Conformal representation of RG-improved f(R) theories\label{sect3}}

Replacing the running couplings in the scale-dependent action $\eqref{eq:qaction}$
yields an effective gravitational action of the form
\begin{equation}
S_{grav}^{\text{{eff}}}=\int d^{4}x\sqrt{-g}\;\left\{ \frac{R^{2}}{128\pi g_{\ast}\lambda_{\ast}}+f_{RG}(R)\right\} \;.\label{eq:actionJF}
\end{equation}
Provided that $f_{RG}^{(2)}(R)\neq-\frac{1}{64\pi g_{\ast}\lambda_{\ast}}$,
and introducing the field $\varphi\equiv16\pi G_{0}\left(\frac{\chi}{64\pi g_{\ast}\lambda_{\ast}}+f'_{RG}(\chi)\right)$,
this action can be re-expressed as
\begin{equation}
S_{grav}^{\text{{eff}}}=\int d^{4}x\sqrt{-g}\;\left\{ \frac{1}{16\pi G_{0}}\varphi R-U(\varphi)\right\} \;,
\end{equation}
where the function $U(\varphi)$ is given by
\begin{equation}
U(\varphi)=\frac{\chi[\varphi]^{2}}{128\pi g_{\ast}\lambda_{\ast}}-f_{RG}(\chi[\varphi])+\chi[\varphi]\;f'_{RG}(\chi[\varphi])\;.
\end{equation}
It is now convenient to perform a conformal transformation, mapping
the metric $g_{\mu\nu}$ in the Jordan frame to the metric $g_{\mu\nu}^{E}=\varphi\,g_{\mu\nu}$
in the Einstein frame. Rescaling the metric $g_{\mu\nu}$ by the conformal factor $\varphi=e^{\sqrt{2/3}\phi/M_{Pl}}$ maps the purely gravitational theory $\eqref{eq:actionJF}$ to General Relativity (i.e. Einstein-Hilbert action) minimally coupled to the scalar field $\phi$ 
\begin{equation}
S_{grav}^{\text{{eff}}}=\int d^{4}x\sqrt{-g_{E}}\;\left(\frac{R_{E}}{16\pi G_{0}}+\frac{1}{2}g_{E}^{\mu\nu}\partial_{\mu}\phi\partial_{\nu}\phi-V(\phi)\right)\;.\label{eq:Eaction}
\end{equation}
The scalar degree of freedom introduced by the function $f_{RG}(R)$
in the Jordan frame, can thus be seen as a scalar field subject to
the potential $V[\varphi(\phi)]=U(\varphi)\,\varphi^{-2}$ in the
Einstein frame.

Due to the coupling to gravity, and depending on the form of the potential $V(\phi)$, the time evolution of the scalar field $\phi$ might lead to a period of inflation. In fact, specializing the metric $g_{\mu\nu}^{E}$ to that of a Friedmann-Lema\i tre-Robertson-Walker (FLRW) universe, the time evolution of the scalar field $\phi(t)$ and the growth of the scale factor $a(t)$ are related to each other and are described
by the following Friedmann and Klein-Gordon equations
\begin{equation}
\left(\frac{\dot{a}}{a}\right)^{2}=\frac{8\pi G_{0}}{3}\left\{ \frac{\dot{\phi}^{2}}{2}+V(\phi)\right\} \;,\label{eq:frieq}
\end{equation}
\begin{equation}
\ddot{\phi}+3H\dot{\phi}+V'(\phi)=0\;.
\end{equation}
In the slow-roll approximation the kinetic energy of the inflaton
field is negligible, $\dot{\phi}^{2}\ll V(\phi)$, and $\ddot{\phi}\ll3H\dot{\phi}+V'(\phi)$.
Note that the reliability of the slow-roll approximation is corroborated by the recent Planck data which, so far, have not found hints for inflationary dynamics beyond slow roll \cite{Akrami:2018odb}.

The first and second variations of the potential define the slow roll
parameters
\begin{equation}
\epsilon(\phi)=\frac{M_{Pl}^{2}}{2}\left(\frac{V'(\phi)}{V(\phi)}\right)^2\;\;,\qquad\eta(\phi)=M_{Pl}^{2}\left(\frac{V''(\phi)}{V(\phi)}\right)\;\;.
\end{equation}
The slow-roll conditions $\epsilon\ll1$ and $\eta\ll1$ are satisfied
when the evolution of the scalar field $\phi$ along its potential
$V(\phi)$ is slow in comparison to the rate of exponential expansion
of the universe. The violation of the slow-roll conditions, encoded
in the equation $\epsilon(\phi_{f})=1$, defines the value of the
field at the end of inflation, $\phi_{f}\equiv\phi(t_{f})$. Fixing
the number of e-folds before the end of inflation
\begin{equation}
N_{e}(\phi_{i})=\int_{\phi_{f}}^{\phi_{i}}\frac{V(\phi)}{V'(\phi)}d\phi\;\;,
\end{equation}
to $N_{e}\simeq60$ \cite{Baumann:2009ds}, provides the initial condition $\phi_{i}\equiv\phi(t_{i})$. The spectral index and tensor-to-scalar ratio 
\begin{equation}
n_{s}=1-6\,\epsilon(\phi_{i})+2\,\eta(\phi_{i})\;\;,\qquad r=16\,\epsilon(\phi_{i})\;\;,\label{eq:nsr}
\end{equation}
are determined by the values of the slow-roll parameters at the
beginning of the period of exponential growth of the universe, i.e., at $\phi=\phi_i$. Therefore, under the slow-roll approximation, the theoretical values $\eqref{eq:nsr}$ can be determined and compared to the values provided by the analysis of the observational data on the anisotropies of the CMB. 

Finally, for a single-field inflationary model with inflationary potential $V(\phi)$, the amplitude of the primordial scalar power spectrum takes the form \cite{Akrami:2018odb}
\begin{equation}\label{amplitude}
A_{s}=\frac{V(\phi_{i})}{24\pi^{2}M_{Pl}^{4}\;\epsilon(\phi_{i})}\simeq2.2\cdot10^{-19}\;.
\end{equation}
Every inflationary model has to be normalized in order to fit this
value (see \cite{inflationaris} for details), and this normalization
fixes the order of magnitude of the inflaton mass to~$m\sim10^{13}\div10^{14}\,\mathrm{GeV}$ \cite{inflationaris}.

\section{The inflationary mechanism in Asymptotically Safe Gravity} \label{sect4}

The RG-improved $f(R)$-type action $S_{grav}^{\text{{eff}}}$, and the corresponding scalar potential $V(\phi)$ in the Einstein frame, depend on the critical exponents $\theta_{i}$. Here we assume that the density fluctuations at the last scattering surface are
generated by the amplification of quantum-gravity fluctuations in
the pre-inflationary era. Hence, requiring the compatibility of
the inflationary dynamics generated by the effective action $S_{grav}^{\text{{eff}}}$ with the Planck data constrains the universality properties of the theory in the vicinity of the NGFP. Moreover, since the gravitational critical exponents $\theta_{i}$ are influenced by the presence of matter~\cite{Dona:2013qba,Biemans:2017zca,Alkofer:2018fxj}, the conditions on the critical exponents imposed by compatibility with observations
could be used, at least in principle, to identify the primordial matter content of universe~\cite{Bonanno:2018gck}. We remark however that the current systematic uncertainties on the computation of the critical exponents are still too large to put strong constraints on the matter content of the early universe based on the aforementioned conditions on the critical exponents.

\begin{figure}[t!]
\centering{}\includegraphics[width=0.9\textwidth]{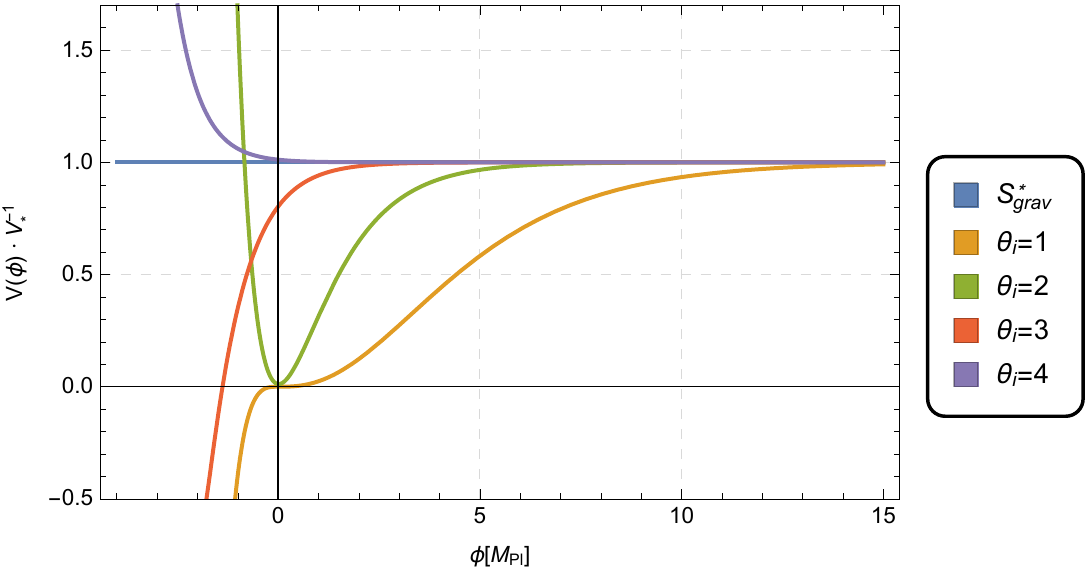}\caption{Inflationary potential $V(\phi)$ generated by the conformal transformation
of the fixed-point action $S_{grav}^{\ast}$, eq. $\eqref{eq:FPaction}$,
and of the effective action $\eqref{eq:effaction}$ for various values
of the critical exponents $\theta_{i}$ in the range $\theta_{i}\in(0,4]$\label{Fig1}.
The fixed-point action $S_{grav}^{\ast}$ gives rise to a flat potential $V_{\ast}=8\pi g_{\ast}\lambda_{\ast}M_{Pl}^{4}$, relict of the fixed-point epoch, and corresponds to a perfectly scale-invariant regime. Moving away from the NGFP,
the renormalization group flow generates additional operators which
destabilize the fixed-point potential $V_{\ast}$. The form of the
modified potential $V(\phi)=V_{\ast}+\delta V(\phi)$ depends on the deviation $\delta\mathcal{L}=f_{RG}(R)$ from the fixed-point regime realized in the Jordan frame. Modeling this variation by means of the RG-improved model $\eqref{eq:qaction}$, the analytical form of the inflationary potential $V(\phi)$ is determined by the critical exponents $\theta_{i}$. Interestingly, the case $\theta_{1}=\theta_{2}=2$ reproduces the well-known Starobinsky model. The departure from the fixed-point regime modifies the inflationary potential such that $V'(\phi)\protect\neq0$ at $\phi\sim M_{Pl}$, and thereby induces a non-zero kinetic energy,
$\dot{\phi}_{i}\sim-V'(\phi_{i})/3H(t_{i})$, for the inflaton field.
This quantity provides an initial boost for the subsequent evolution
of the scale factor $a(t)$, i.e. for $t>t_{i}$, according to eq.
$\eqref{eq:frieq}$.}
\end{figure}

The fixed-point regime is described by the action $S_{grav}^{\ast}$. Following the procedure described in the previous section, it is not
difficult to see that the fixed-point action $S_{grav}^{\ast}$ is
conformally equivalent to a scalar-tensor theory $\eqref{eq:qaction}$, where the scalar field $\phi$ is minimally coupled to gravity and subject to the constant potential
\begin{equation}
V_{*}(\phi)=8\pi g_{\ast}\lambda_{\ast}M_{Pl}^{4}\;.\label{eq:flatpotential}
\end{equation}
The fixed-point potential $V_{\ast}$ should give rise to a perfectly
scale-invariant scalar power spectrum. However, as the RG flow moves away from the NGFP, additional operators are generated and the potential $V_{\ast}$ is dynamically modified by this running
\begin{equation}
V_{\ast}\to V(\phi)=V_{\ast}+\delta V(\phi)\;.
\end{equation}
The deviation $\delta V(\phi)$ of the scalar potential from its fixed-point value $V_{\ast}$ is determined by the function~$\delta\mathcal{L}=f_{RG}(R)$ and, in accordance with the simple RG-improved model introduced above, it depends on the critical exponents $\theta_{i}$. Its analytical form can be determined by performing a conformal transformation of the original $f_{RG}(R)$ theory, as detailed in sect. \ref{sect3}. As already mentioned, we assume the critical exponents to be real numbers.
The results are displayed in Fig. $\text{\ref{Fig1}}$ for the special
case $\theta_{1}=\theta_{2}$, and for $\theta_{i}$ taking values
$0<\theta_{i}\leq4$. These bounds are justified as follows:
\begin{itemize}
\item The critical exponents are assumed to be positive, $\theta_{i}>0$,
in order to fulfill the asymptotic-safety condition;
\item There exists at least one critical exponent $\theta_{i}<4$. As is
clear from the form of the RG-improved action $\eqref{eq:qaction}$,
this condition ensures that the scalar power spectrum deviates from the perfect scale-invariance realized by the fixed-point
regime, and thereby guarantees compatibility with the Planck data
(see also \cite{Bonanno:2018gck} for details). The borderline case
$\theta_{i}=4$ works under very specific assumptions, and will be
discussed in detail below.
\end{itemize}

Note that, due to the structure of the function $f_{RG}(R)$ and provided that $\theta_{i}\neq0$, the $R^{2}$ term in the action $\eqref{eq:qaction}$ does not gain any additional contribution from the operators in $\delta\mathcal{L}=f_{RG}(R)$ (at least in the  regime where the couplings scale as in eq.~\eqref{eq:scalingUV}).
Therefore, its coefficient inherits the universality properties of
the fixed-point action $S_{grav}^{\ast}$ and defines a mass scale
\begin{equation}
m^{2}=8\pi\left(\frac{4}{3}\,\lambda_{\ast}g_{\ast}\right)M_{Pl}^{2}\label{eq:infmass}
\end{equation}
determined by the universal product $\lambda_{\ast}g_{\ast}$ \cite{Lauscher:2001ya}, which is typically $\sim\mathcal{O}(1)$. It is interesting to notice that when higher derivative operators
\begin{equation}
\mathcal{L}_\mathrm{HD}=\frac{1}{16\pi G_k}\sum_{n\geq2} \frac{\zeta^{(n)}_k}{n}\frac{R^n}{(3 k^2)^{n-1}}\;\; 
\end{equation}
are included in the ansatz for the bare action, $\zeta^{(n)}$ being dimensionless couplings, the RG-improvement generated by the consistency equation \eqref{eq:cutid} preserves the structure of the fixed-point effective action. Specifically, $S_{grav}^\ast$ will maintain its pure $R^2$-form, while the mass scale $m$ will be corrected by the presence of other fixed-point quantities. Assuming that a UV-attractive NGFP persists in arbitrary large truncations, the new fixed-point potential $V_\ast$ yields the mass scale
\begin{equation} \label{planckic}
m^{2}=8\pi\left(\frac{\frac{4}{3}\,\lambda_{\ast}g_{\ast}}{1+\sum_{n\geq2}\frac{2}{n}\big(\frac{4 \lambda_\ast}{3}\big)^{n-1}\,\zeta^{(n)}_\ast}\right)M_{Pl}^{2}\;\;.
\end{equation}
The general form of the RG-improved action~\eqref{eq:effaction} is preserved as well, but a contribution from the critical exponent $\theta_3$ due to the additional relevant operator $R^2$ should appear.
The normalization of the scalar CMB power spectrum dictated by eq.~\eqref{amplitude} implies a non-trivial constraint on the fixed-point values of the dimensionless couplings $(g_k,\lambda_k,\zeta^{(n)}_k)$, 
\begin{equation}
\left(\frac{\frac{4}{3}\,\lambda_{\ast}g_{\ast}}{1+\sum_{n\geq2}\frac{2}{n}\big(\frac{4 \lambda_\ast}{3}\big)^{n-1}\,\zeta^{(n)}_\ast}\right)\simeq10^{-6}\;\;.
\end{equation}
As functional RG computations indicate that $\lambda_{\ast}g_{\ast}\sim\mathcal{O}(1)$ \cite{Lauscher:2001ya}, the sum in the denominator would be required to be very large in order to fit observations. On the other hand, going beyond the fixed-point approximation introduced in sect. \ref{sect2} modifies the scaling relations~\eqref{eq:scalingUV}. If the integration of fast-fluctuating modes generates additional contributions to the $R^2$-operator (or, equivalently, if the coupling to the $R^2$ operator runs), the ``plateau scale''~\eqref{planckic} will be dynamically modified by the departure of the flow from the fixed-point regime. In this case the initial conditions for inflation would be defined by the fixed-point mass scale \eqref{planckic}, that can even be Planckian, while the amplitude of the scalar fluctuations at the horizon exit, eq.~\eqref{amplitude}, would be fixed by the ``renormalized'' plateau scale. If this dynamical-plateau mechanism is realized, it could provide a solution to the ``unlikeness problem'' raised in \cite{Ijjas:2013vea}. As addressing this question goes far beyond the purpose of these proceedings, in what follows we will restrict ourselves to the fixed-point approximation introduced in sect. \ref{sect2}.

The departure from the fixed-point regime causes an instability of
$V_{\ast}$, and results in a scalar potential $V(\phi)$ whose first
derivative is non-zero at $\phi_{i}\sim M_{Pl}$. As a consequence,
even if $\dot{\phi}=0$ in the NGFP era, the inflaton field acquires
a non-zero kinetic energy, $\dot{\phi}_{i}\sim-V'(\phi_{i})/3H(t_{i})$, which provides an initial boost towards the subsequent time evolution of the universe. In particular, depending on the shape of the potential, this mechanism can potentially trigger a period of slow-roll inflation, the growth of the scale factor $a(t)$ being controlled by the function
$V(\phi(t))$ according to eq. $\eqref{eq:frieq}$. Particularly interesting
is the case $\theta_{1}=\theta_{2}=2$, which realizes a Starobinsky-like potential (see Fig. $\text{\ref{Fig1}}$)
\begin{equation}
V(\phi)=e^{-2\sqrt{\frac{2}{3}}\frac{\phi}{M_{Pl}}}\left\{ \frac{3}{4}\left(1-e^{\sqrt{\frac{2}{3}}\frac{\phi}{M_{Pl}}}\right)^{2}m^{2}+\Lambda_{\mathrm{{eff}}}\right\} M_{Pl}^{2}\;,\label{eq:theta2}
\end{equation}
in the presence of an effective cosmological constant $\Lambda_{\mathrm{{eff}}}=-(b_{1}+b_{4}+b_{5})M_{Pl}^{2}$.
As it is well known, this model leads to cosmic parameters
\begin{equation}
n_{s}\simeq1-\frac{2}{N_{e}}+\mathcal{O}(N_{e}^{-3})\;,\qquad r\simeq\frac{12}{N_{e}^{2}}+\mathcal{O}(N_{e}^{-3})\;,
\end{equation}
in good agreement with the Planck data. The case $\theta_{1}=\theta_{2}=4$
is a limiting case where the function $f_{RG}(R)$ reduces to a constant
term $\Lambda_{\mathrm{{eff}}}=-(b_{2}+b_{3})M_{Pl}^{2}$, plus an additional $R^{-2}$ operator. Operators of the form $R^{-\alpha}$,
with $\alpha>0$, are suppressed at curvature scales $R\gtrsim M_{Pl}^{2}$ and hence do not contribute to the inflationary dynamics \cite{Bonanno:2018gck}. Overall, neglecting the $R^{-2}$ contribution to the action $\eqref{eq:qaction}$,
the inflationary potential can be approximated by
\begin{equation}
V(\phi)=V_{\ast}+e^{-2\sqrt{\frac{2}{3}}\frac{\phi}{M_{Pl}}}\Lambda_{\mathrm{{eff}}}M_{Pl}^{2}\;.\label{eq:theta4}
\end{equation}
This potential has no local minima and, depending on the sign of $\Lambda_{\mathrm{{eff}}}$,
diverges to $\pm\infty$ as $\phi\to-\infty$. As a consequence, no
reheating phase by standard parametric-oscillations of the inflaton
field is possible, and a new mechanism to reheating the universe after
inflation would be required. The spectral index and tensor-to-scalar
ratio generated by the class of potentials $\eqref{eq:theta4}$ read
\begin{equation}
n_{s}\simeq1+\frac{3\Lambda_{\mathrm{{eff}}}}{8\pi(g_{\ast}\lambda_{\ast})M_{Pl}^{2}N_{e}^{2}}+\mathcal{O}(N_{e}^{-3})\;,\qquad r\simeq\frac{27\Lambda_{\mathrm{{eff}}}^{2}}{256\pi^{2}(g_{\ast}\lambda_{\ast})^{2}M_{Pl}^{4}N_{e}^{4}}+\mathcal{O}(N_{e}^{-5})\;.
\end{equation}
Assuming $g_{\ast}\lambda_{\ast}>0$, these numbers are compatible
with observations if $\Lambda_{\mathrm{{eff}}}$ is sufficiently small
and negative. Note that in principle the presence of a negative $\Lambda_{\mathrm{{eff}}}$
at inflationary scales is not necessarily incompatible with the current
phase of accelerated expansion of the universe. In fact, for~$k^{2}\lesssim M_{Pl}^{2}$ the approximation~$\eqref{eq:scalingUV}$ breaks down and additional operators start contributing to the scale-dependent action~$\eqref{eq:qaction}$.
In particular, operators of the form~$R^{-\alpha}$ start playing
a role at cosmological scales and could overcome the effects of a
negative cosmological constant, and drive the late-time evolution
of the universe towards the current phase of accelerated expansion
\cite{defe}. The case of a positive effective cosmological constant,
with $g_{\ast}\lambda_{\ast}>0$, would instead lead to a spectral
index $n_{s}\gtrsim1$ and would not allow for a ``natural'' exit
from the inflationary phase by violation of the slow-roll conditions.
In this case compatibility with the Planck data would rather constrain
the ultraviolet value of the cosmological constant to be negative,
$\lambda_{\ast}<0$ (see also \cite{Bonanno:2017gji,Eichhorn:2017ylw,Biemans:2017zca,Bonanno:2018gck}). Within the present fixed-point approximation, where the inflaton mass is defined by eq. \eqref{eq:infmass}, a negative ultraviolet cosmological constant would entail the presence of a tachionic inflaton field. This result seems in contradiction with \cite{Bonanno:2018gck},
where the \emph{avoidance} of a tachionic inflaton field required
$\lambda_{\ast}<0$. This mismatch is caused by a different ratio
$k^{2}/R$: in \cite{Bonanno:2018gck} the infrared cutoff $k$ was
related to the Ricci scalar by means of an unspecified positive constant $\xi$, $k^{2}=\xi\,R$. The inflaton mass is thus given
by $m^{2}=\frac{g_{\ast}(8\pi M_{Pl}^{2})}{6\xi(1-2\xi\lambda_{\ast})}$, and is positive for arbitrary values of $\xi$ only when $\lambda_{\ast}<0$. However, if $\lambda_{\ast}$ sets the scale of Quantum Gravity according to eq. $\eqref{eq:cutid}$, then $\xi(\lambda_{\ast})\equiv1/4\lambda_{\ast}$ is no longer an arbitrary constant, and our result $\eqref{eq:infmass}$
is recovered.

\section{Conclusions} \label{sect5}

According to the Asymptotic Safety conjecture, in the deep ultraviolet  the gravitational interaction reaches a scale-invariant regime due to the existence of a non-trivial fixed point of the renormalization group (RG) flow. In these proceedings we have reviewed some key consequences of the existence of a fixed-point epoch in the ``RG-improved'' cosmological evolution of the universe. 

Assuming that the running of the gravitational couplings can be incorporated into the classical spacetime dynamics by means of a scale-dependent infrared cutoff $k(x)$, a period of slow-roll inflation can be associated with the displacement of the RG flow from the regime where the theory is scale invariant. Following \cite{2004reuterw1,Domazet:2012tw,Koch:2014joa}, the invariance of the theory under diffeomorphisms requires the scale-dependent cutoff~$k(x)$ to vary at the same rate as the scalar curvature $R$.
Starting from a scale-dependent Einstein-Hilbert action, the replacement~$k^2\to R$ generates an effective $f(R)$-Lagrangian of the form $\mathcal{L}_{grav}^{\text{{eff}}}=\mathcal{L}_{\ast}+\delta\mathcal{L}_{RG}$, where $\mathcal{L}_{\ast}\sim R^2$ is realized at the fixed point, in accordance with several FRG computations \cite{Benedetti:2012dx,Dietz:2012ic,Demmel:2015oqa}, and $\delta\mathcal{L}_{RG}$ introduces additional operators, mimicking the effect of a Wilsonian RG flow: although starting from a simple bare theory at the fixed point, at lower energy scales the effective degrees of freedom interact through more complicated interactions, generated by the integration of high-frequency fluctuating modes in the functional integral. This mechanism could provide a simple explanation for the origin and distribution of the temperature fluctuations of the Cosmic Microwave Background (CMB). The fixed point action $S_{grav}^\ast$ describes in fact a scale-invariant gravitational theory and, in the Einstein frame, corresponds to Einstein gravity coupled to a scalar degree of freedom whose interaction potential is constant, $V_\ast=8\pi g_\ast\lambda_\ast M_{Pl}^4$. This potential defines the universal mass scale~$m^2\sim (\lambda_\ast g_\ast)M_{Pl}^2$.
The departure from the fixed-point regime destabilizes the fixed-point potential~$V_\ast\to V(\phi)=V_\ast+\delta V(\phi)$, the correction~$\delta V(\phi)$ corresponding to the variation $\delta\mathcal{L}_{RG}$ of the gravitational Lagrangian caused by the Wilsonian RG flow. Specifically, for~$\phi\lesssim M_{Pl}$, the scalar potential develops a minimum or diverges. The region~$\phi\gg M_{Pl}$ remains instead unaffected: the renormalized scalar potential $V(\phi)$ is characterized by a plateau region, where $V_{plateau}(\phi)=m^2 M_{Pl}^2$, relict of the fixed-point epoch. 
The nearly-scale invariant scalar power spectrum is thus understood as the result of the nearly-scale-invariant behavior of gravity in the vicinity of the ultraviolet fixed point.

In the RG-improved model introduced in \cite{Bonanno:2018gck} and revisited here, the RG running of the couplings is approximated by their scaling about the non-trivial fixed point. Due to this approximation, the variation $\delta\mathcal{L}_{RG}$ does not yield additional $R^2$ operators, and leaves the mass scale $m^2\sim (\lambda_\ast g_\ast)M_{Pl}^2$ unaffected. In a more complete description, accounting for the full running of the couplings and possibly including higher-derivatives operators in the action, we expect the coupling of the quadratic term $R^2$ to run and to redefine the plateau scale~\cite{AlfioAle1}. The scale of inflation is then determined by the renormalized plateau. This dynamical-plateau scenario would allow to setup the initial conditions for inflation at Planckian energies, where the scalar potential is constant and $m^2\propto (\lambda_\ast g_\ast)M_{Pl}^2$, while being able to reproduce the correct amplitude of scalar perturbations at the horizon exit. This mechanism could provide a solution to the ``unlikeness problem'' raised in~\cite{Ijjas:2013vea}. However, this behavior is not captured by the simple RG improved model reviewed here and addressing this problem requires to go beyond the fixed-point approximation employed in the present proceedings.

The departure of the RG flow from the fixed-point regime induces a scalar potential $V(\phi)$ characterized by a plateau and can potentially trigger a period of slow-roll inflation. The specific form of the potential $V(\phi)$ depends on universality properties of the gravitational RG flow and, as a consequence, a comparison with the observational data can put constraints on the gravitational critical exponents \cite{Bonanno:2018gck}. Specifically, since the power spectrum of primordial scalar perturbations is almost scale-invariant  but not exactly scale-invariant, at least one of the critical exponents must be $\theta_{i}<4$. Among the class of inflationary models derived from the action $\eqref{eq:qaction}$, the case $\theta_{1}=\theta_{2}=2$ reproduces the well-known Starobinsky model and is thereby compatible with observations.

It would be interesting to understand if similar conclusions can be drawn by using a self-consistent RG-improvement \cite{Platania:2019kyx} of the classical cosmological solutions. In a first approximation, it has been shown that the anti-screening character of the Newton coupling could replace the classical initial singularity with a regular bounce or with an emergent universe scenario \cite{Bonanno:2017gji}, both characterized by a non-vanishing minimum value of the scale factor and a period of inflation following the bounce. Finding the class of actions giving rise to this type of regular cosmologies is the first step towards understanding the relation between the results obtained in \cite{Bonanno:2016rpx,Bonanno:2018gck} and summarized in these proceedings, and the inflationary scenario following a cosmological bounce \cite{Bonanno:2017gji}. We reserve to discuss these problems in future works. 

\begin{acknowledgments}
{I thank the organizers of the workshop ``Quantum Fields — From Fundamental Concepts to Phenomenological Questions'' for creating a highly stimulating scientific atmosphere and all participants for interesting discussions. I also thank A. Eichhorn for discussions and comments on this manuscript. The research of AP is supported by the Alexander von Humboldt Foundation.}
\end{acknowledgments}

\bibliography{AleBib}

\end{document}